\documentclass{aa}
\newcommand{\msun}{$M_{\odot}$}
\newcommand{\teff}{$T_{\rm eff}$}
\newcommand{\lsiv}{\object{LS\,IV$-$14$\degr$116}}
\newcommand{\feige}{\object{Feige\,46}}
\newcommand{\kms}{km\,s$^{-1}$}
\usepackage{natbib}
\usepackage{graphicx}
\usepackage{txfonts}
\usepackage{color}
\usepackage{booktabs}
\usepackage[dvipsnames]{xcolor}

\begin{document}

\title{Heavy-metal enrichment in the intermediate He-sdOB pulsator \feige }

\author{ M. Latour\inst{1}, M. Dorsch\inst{2,3}, and U. Heber\inst{3} }  

\institute{
Institute for Astrophysics, Georg-August-University, 
Friedrich-Hund-Platz 1, 37077 G\"{o}ttingen, Germany \\
  \email{marilyn.latour@uni-goettingen.de } 
\and
Institut für Physik und Astronomie, Universität Potsdam, Haus 28, Karl-Liebknecht-Str. 24/25, 14476 Potsdam-Golm, Germany
\and
Dr. Karl Remeis-Observatory \& ECAP, Friedrich-Alexander University Erlangen-N\"{u}rnberg,
Sternwartstr. 7, 96049 Bamberg, Germany
  }

\date{Received ; accepted }

\abstract
{The intermediate He-enriched hot subdwarf star \feige\ was recently reported as the second member of the V366 Aqr (or He-sdOBV) pulsating class. \feige\ is very similar to the prototype of the class, \lsiv, not only in terms of pulsational properties, but also in terms of atmospheric parameters and kinematic properties. \lsiv\ is additionally characterized by a very peculiar chemical composition, with extreme overabundances of the trans-iron elements Ge, Sr, Y, and Zr. In this paper, we investigate the possibility that the similitude between both pulsators extends to their chemical composition. We retrieved archived optical and UV spectroscopic observations of \feige\ and perform an abundance analysis using model atmospheres and synthetic spectra computed with \texttt{TLUSTY} and \texttt{SYNSPEC}.
In total, we derive abundances for 16 metallic elements and provide upper limits for four additional elements. 
From absorption lines in the optical spectrum of the star we measure an enrichment of more than 10\,000$\times$ solar for yttrium and zirconium. As for strontium, the UV spectrum revealed it to be equally enriched. Our results confirm that \feige\ is not only a member of the now growing group of ``heavy-metal'' subdwarfs, but also has an abundance pattern remarkably similar to that of \lsiv. 
}


 \authorrunning{Latour, Dorsch \& Heber}
 \titlerunning{Heavy-metal enrichment in \feige}

\maketitle

\section{Introduction}

Hot subdwarf stars of spectral type B and O (sdB and sdO) are low-mass ($\sim$0.5\msun) and evolved objects. There is a consensus that sdB stars populate the extreme horizontal branch (EHB), which is the very hot end (\teff\ > 20 000 K) of the horizontal branch. 
The helium-rich sdO stars, however, are too hot and luminous to be EHB objects. Whether or not their evolution is tied to the EHB is still under debate.
The vast majority of hot subdwarfs produce energy via helium burning, either in the core or as shell-burning in the most evolved objects.  
The extreme lightness of their hydrogen envelope prevents hot subdwarf stars from ascending the asymptotic giant branch and, instead, they evolve directly to the white dwarf cooling sequence once the core He-burning ceases \citep{dor93,heb16}.

The relatively high surface gravity (4.8 $\la$ log $g$ $\la$ 6.4) of hot subdwarfs, combined with the high temperature of their radiative atmosphere, make the chemical composition of these stars strongly affected by diffusion effects. Gravitational settling and radiative levitation are the main phenomena altering the surface composition, but stellar winds and turbulence are also believed to contribute (see, e.g., \citealt{ung08,mich11,hu11}). A consequence of the diffusion process is that all hot subdwarfs are chemically peculiar. 
One of their chemical particularities is their wide variety of helium abundances, ranging from helium dominated atmospheres in the extremely He-rich (eHe) sdOs, with helium-to-hydrogen number density ratio (log $N$(He)/$N$(H)) up to 3, to the helium-depleted atmospheres present in the majority of the sdBs, with log $N$(He)/$N$(H) down to $-$4 \citep{heb16}. While the variety of helium abundances found among the He-poor sdB stars are likely to be the result of diffusion processes, the helium enrichment observed in He-sdOs is best explained by mixing related to their non-canonical evolutionary history \citep{bro01,bert11,zhang12}. 

A relatively small fraction of hot subdwarfs have atmospheres that are moderately enriched in helium ($-1.0$ $\la$ log $N$(He)/$N$(H) $\la$ 0.6) and are referred to as intermediate helium (iHe) subdwarfs \citep{nas13,geier17}. Such mild helium enrichment is mostly found in stars that are close to the transition between the sdB and sdO spectral types (i.\,e.~with \teff\ $\sim$ 35 000$-$40 000 K). Stars in this temperature range are often attributed an sdOB spectral type, as they show a \ion{He}{ii} 4686 \AA\ line in addition to \ion{He}{i} lines.
This small class of hot subdwarfs would go rather unnoticed, and has been until recently, if some of them were not to display extreme enrichment in heavy and trans-iron elements. 
This was first noticed by \citet{nas11} with the identification of yttrium, zirconium and strontium lines in the optical spectrum of \lsiv\ that were associated with overabundances of about 4 dex with respect to solar values. Since then a few more objects have joined the group of ``heavy-metal'' stars (\citealt{nas13,jeff17}), with some also displaying significant overabundances of lead. Most recently, Dorsch et al. (2019; submitted) added two stars, HZ\,44 and HD\,127493, to the class of heavy-metal rich iHe subdwarfs. The spectral analysis of HZ\,44 is the most extensive so far, as it provided elemental abundances for 29 species, including trans-iron elements, and upper limits for the abundances of 10 additional elements.
Although overabundances of heavy elements are observed in some hydrogen-rich sdBs as well \citep{otoole06,chayer2006,bla08}, their abundances remain smaller than in the heavy-metal iHe subdwarfs.

Apart from its abundance peculiarities, \lsiv\ is also known for its puzzling pulsations properties. It is the prototype of the very small class of V366 Aqr (or He-sdOBV) pulsators, for which the driving mechanism remains hypothetical \citep{saio19,bat18,bert11}. This class contains so far only two members: \lsiv\ itself and \feige. Besides their pulsation properties, both stars also share similar atmospheric properties (\teff, $\log g$ and He abundance) and kinematics typical of the galactic halo population \citep{ran15,lat19}. As opposed to \lsiv, very little is known about the chemical composition of \feige. Only the work of \citet{bauer95} provided C and N abundance estimates close to solar, while their upper limits on Mg, Si and Al indicated that these elements are depleted.

With \feige\ being an iHe-sdOB (with log $N$(He)/$N$(H) = $-$0.3) and otherwise so similar to \lsiv, 
we decided to further investigate its chemical composition and look for these heavy elements that are so conspicuous in the atmosphere of its pulsating sibling. 
In Sect.~2 we present the spectroscopic observations and model atmospheres used to perform the abundance analysis, which is described in Sect.~3. We make use of the parallax and magnitude measurements of \feige\ to derive its mass, radius and luminosity in Sect.~4. Finally, a discussion and short conclusion are presented in Sect.~5 and 6. 

\section{Methods}
Two available spectral observations have a sufficient resolution to be used for an abundance analysis.
The first one is an optical spectrum obtained by one of us (U.~H., \citealt{dri87}) with the Cassegrain Echelle Spectrograph (CASPEC) formerly installed at the ESO 3.6-metre telescope at La Silla Observatory in Chile. The data was obtained on April 4, 1984 with a total exposure time of 90 min. The spectrum covers the 3860$-$4840 \AA\ range at a resolution of R~$\approx$~20 000 and
has a signal-to-noise ratio (SNR) of about 45. We note that this is the same spectrum used by \citet{bauer95}. 
In addition to some optical coverage, high-resolution UV observations of \feige\ are available. This star is one of the few iHe-sdOB, and the only iHe pulsator, for which such observations exist. \feige\ has been observed with the Goddard high-resolution spectrograph (GHRS) over the 1323$-$1518 \AA\ range but lacks coverage in the 1359.1$-$1377.4, 1414.1$-$1438.0, and 1474.5$-$1482.5 \AA\ regions. The G160M grating provided a resolution of $\Delta \lambda =$ 0.07 \AA. \feige\ was also observed with the Internation Ultraviolet Explorer (IUE) short wavelength spectrograph using the high-dispersion mode (data ID SWP17466), however the SNR is rather low. Nevertheless, we used this spectrum to look for specific elements that did not show spectral lines in the GHRS range.

The abundance analysis was performed using the method described in \citet{dorsch19}
that was applied to two iHe-sdO stars. In short, non-local thermodynamic equilibrium (NLTE) model atmospheres and synthetic spectra were computed using the \texttt{TLUSTY} and \texttt{SYNSPEC} codes \citep{hubeny88,Lanz2003,hub11}.
As for the atomic data necessary to model the spectral lines, we used the recent Kurucz line list\footnote{http://kurucz.harvard.edu/linelists/gfnew/gfall08oct17.dat} that we supplemented with data collected from various literature sources (such as \citealt{morton00,TOSS15,vanHoof2018}) for elements heavier than zinc. A detailed description of the atomic transitions included in the spectra can be found in \citet{dorsch19}. 
The model atmospheres were computed with the parameters derived by \citet{lat19} (\teff\ = 36100 K, log $g$ = 5.93). As a consistency check, we also fitted selected hydrogen and helium lines available in the CASPEC spectrum with the NLTE model grid computed by \citet{stro07}, using the T\"ubigen Model Atmosphere Package (TMAP) code \citep{tmap12}, including H and He only. We obtained similar atmospheric parameters (\teff\ = 36 200 $\pm$ 1500, log $g$ = 6.1 $\pm$ 0.3) as well as a helium abundance of log $N$(He)/$N$(H) = $-$0.36 $\pm$ 0.07, which is in excellent agreement with the value of $-$0.32 $\pm$ 0.03 obtained by \citet{lat19}.
Nevertheless, since the CASPEC spectrum suffers from normalization deficiencies, we preferred to adopt the atmospheric parameters available from the literature.

As initial abundances for the metallic elements, we used the values reported by \citet{bauer95} for \feige\ and those reported by \cite{nas11} in \lsiv\ for additional elements. Based on this preliminary model atmosphere, series of synthetic spectra with a range of abundances for each element investigated were computed with \texttt{SYNSPEC}. The final model atmosphere includes the following elements and ions in non-local thermodynamic equilibrium (NLTE): \ion{H}{i}, \ion{He}{i}-\textsc{ii}, \ion{C}{ii}-\textsc{iv}, \ion{N}{ii}-\textsc{v}, \ion{O}{ii}-\textsc{vi}, \ion{Mg}{ii}-\textsc{v}, \ion{Si}{ii}-\textsc{iv}, \ion{Fe}{ii}-\textsc{v}, \ion{Ni}{iii}-\text{v}, as well as an additional ion of the following ionization stage represented by its ground state only. The model atoms used are available with the \texttt{TLUSTY} package except for \hbox{Mg~\textsc{iii}-\textsc{v}}. For these ions, we used the same model atoms as in \citet{lat13}. 
Other elements for which we derived abundances are included in the spectral synthesis using the LTE approximation. Details on the calculation of the partition function for trans-iron elements are included in the Appendix A.

 \section{Spectroscopic Analysis}
 
Whenever possible, we fitted spectral lines using the downhill-simplex fitting program \texttt{SPAS} developed by \citet{hirsch09}. However, this method can only be used when spectral lines are relatively well isolated such as those seen in the optical spectrum. Line blending becomes an issue in the UV range and the abundances of some elements showing lines only in the GHRS spectra were estimated by manually comparing models with the observations. 

 \begin{figure*}
 	\sidecaption
 	\includegraphics[width=12cm]{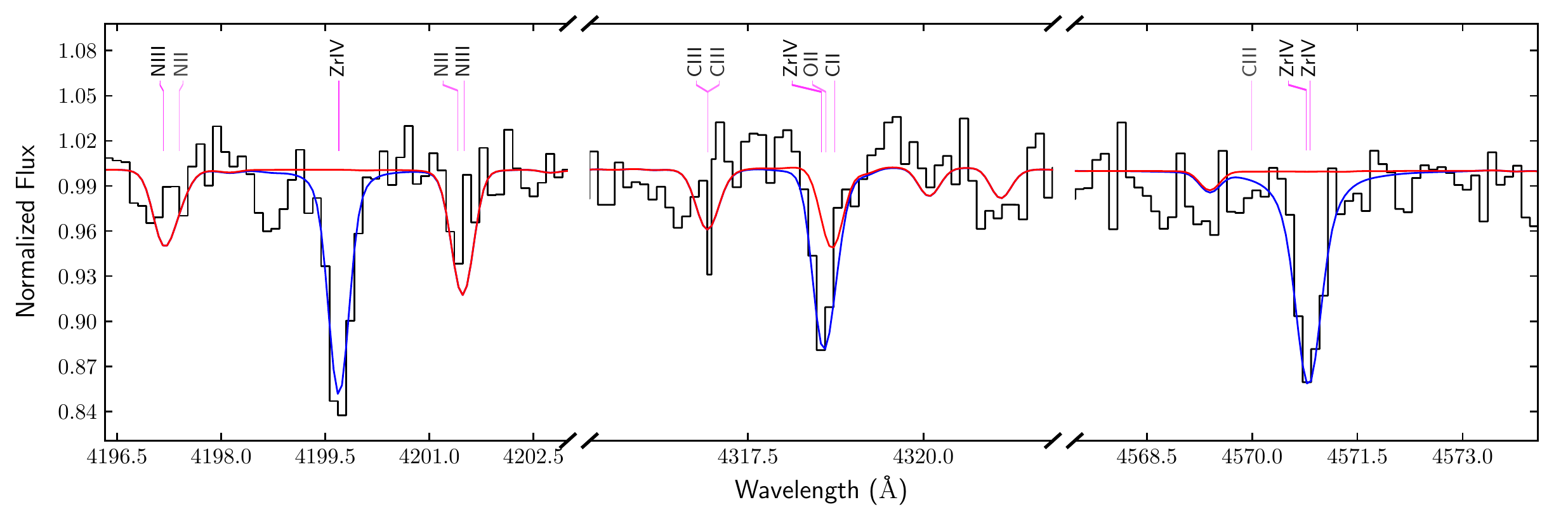}
	\caption{Best fit of the three \ion{Zr}{iv} lines seen in the CASPEC spectrum with a resulting abundance of log Zr/H = $-$5.0. A synthetic spectrum computed without the contribution of Zr is shown as comparison in red. }
 	\label{caspec_zr}
 \end{figure*}
 
We realised during the comparison and fitting of the first metallic lines that additional broadening was needed to reproduce the shape of the spectral lines in the GHRS data. A simultaneous fit of many sharp lines across the GHRS spectra indicated that a rotational broadening of $v \sin i$ $\approx$ 10$-$15 \kms\ provided an optimal reproduction of the line shape. We will further discuss this result in Sect. 5. We included a rotational broadening of $v \sin i$ = 10 \kms\ in our synthetic spectra for the abundance determination.
As for the radial velocity, we obtained from the optical spectrum a value of $90\pm4$ \kms, as reported in \cite{dri87} who used the same spectrum. 

In the following subsections, we describe the abundance analysis of the CASPEC spectrum and the additional abundances we determined using the UV spectra. In the text, we use the shorter notation "log X/H" to refer to the abundance by number with respect to hydrogen "log $N$(X)/$N$(H)".  Our final abundances are presented in Table~\ref{table_abund} where we additionally state the abundances by number fractions ($\epsilon$) where 
$\epsilon_X$~=~$N$(X)~/~$\sum_i N(i)$. The uncertainties are usually computed using the standard deviation of the values obtained from fitting the individual lines. When only few lines are visible in the UV spectra, uncertainties are evaluated by eye (see for example the case of Sr in Sect. 3.2).  
When deriving upper limits, we also include in Table~\ref{table_abund} an uncertainty on that value, which represents a more conservative upper limit. We select the upper limit as the abundance that better matches the observations, while the more "conservative" upper limit results in predicted features that are too strong compared to the observations (see for example Pb in Fig.~\ref{spec_iue}).

\subsection{The optical spectrum} 

Although the SNR of the CASPEC spectrum is not very high, a fair amount of metallic lines are visible. The vast majority of these lines originates from carbon, nitrogen, and oxygen transitions. The strongest lines were individually fitted and we
found C and N to be slightly enriched with respect to the solar values while O is depleted by a factor of ten.
The abundances obtained from the optical lines also reproduce well the C, N, and O lines present in the UV range. The only exception is the \ion{C}{iii} line at 1329.2 \AA\ that is too strong in the model, although other UV lines of \ion{C}{iii} are properly reproduced.
 The Mg abundance derived from the weak Mg\,\textsc{ii}\,$\lambda$4481 line is slightly sub-solar. Because there is no other Mg line visible, and the detected line is weak, we adopt this abundance as an upper limit.
 
The CASPEC spectrum of \feige\ also features three relatively strong \ion{Zr}{iv} lines. Zirconium lines have only been identified so far in the optical spectra of three other hot subdwarfs, all of them belonging to the intermediate helium-rich spectral class (\citealt{nas11,nas13}, Dorsch et al. 2019, submitted). The fit of the three Zr lines resulted in an abundance of log Zr/H = $-$5.0, which translates to an enrichment of about 20\,000 with respect to the solar abundance. 
The best fit to the Zr lines is presented in Fig.~\ref{caspec_zr}. The Zr abundance derived from the optical lines is further supported by the two lines present in the GHRS range (\ion{Zr}{iv} $\lambda\lambda$1441, 1469.5). 

The Y\,\textsc{iii} lines ($\lambda\lambda$4039.602, 4040.112) are also visible
although the quality of the spectrum in that region is not very good. 
The lines are well reproduced with an abundance of log~Y/H~$=~-$5.0.
Unfortunately, these spectral lines are the only yttrium features identified so far in hot subdwarfs. Oscillator strengths for additional yttrium lines, particularly \ion{Y}{iv}, or a higher quality optical spectrum of \feige\ would be helpful to confirm our abundance value.  
 Nevertheless, the abundances derived for both Zr and Y suggest an enrichment similar to that seen in \lsiv. A comparison between the full CASPEC spectrum and our final model is shown in Fig.~\ref{caspec_full}.

\subsection{The UV spectra}
Although the abundances that can be reliably derived with the optical spectrum are limited to a few elements, the UV range covered by the GHRS spectra, and supplemented by the IUE spectrum, allowed us to derive abundances of eleven additional elements. The comparison between the UV spectra and our final model of \feige\ are shown in Figs.~\ref{ghrs_full} and \ref{iue_full} while the values derived for all atomic species are listed in Table~\ref{table_abund}. 
Below, we summarize the chemical composition obtained from the UV observations. 

 \begin{figure*}
 	\centering
 	\includegraphics[width=0.98\textwidth]{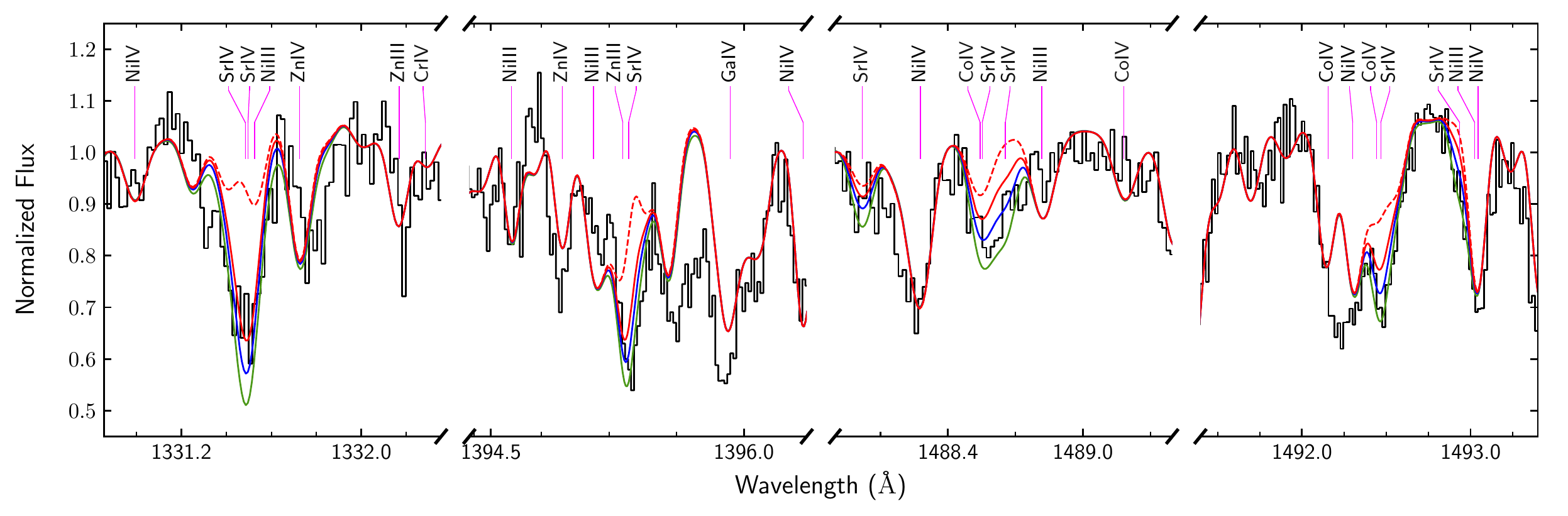}
	\caption{Some of the strongest \ion{Sr}{iv} lines in the GHRS spectra compared with synthetic spectra including [Sr/H] = $-$4.0 (green), $-$4.4 (blue), $-$4.8 (red) and without the contribution of Sr (dashed red). 
	}
 	\label{ghrs_sr}
 \end{figure*}

   \begin{figure*}
 	\centering
 	\includegraphics[width=0.98\textwidth]{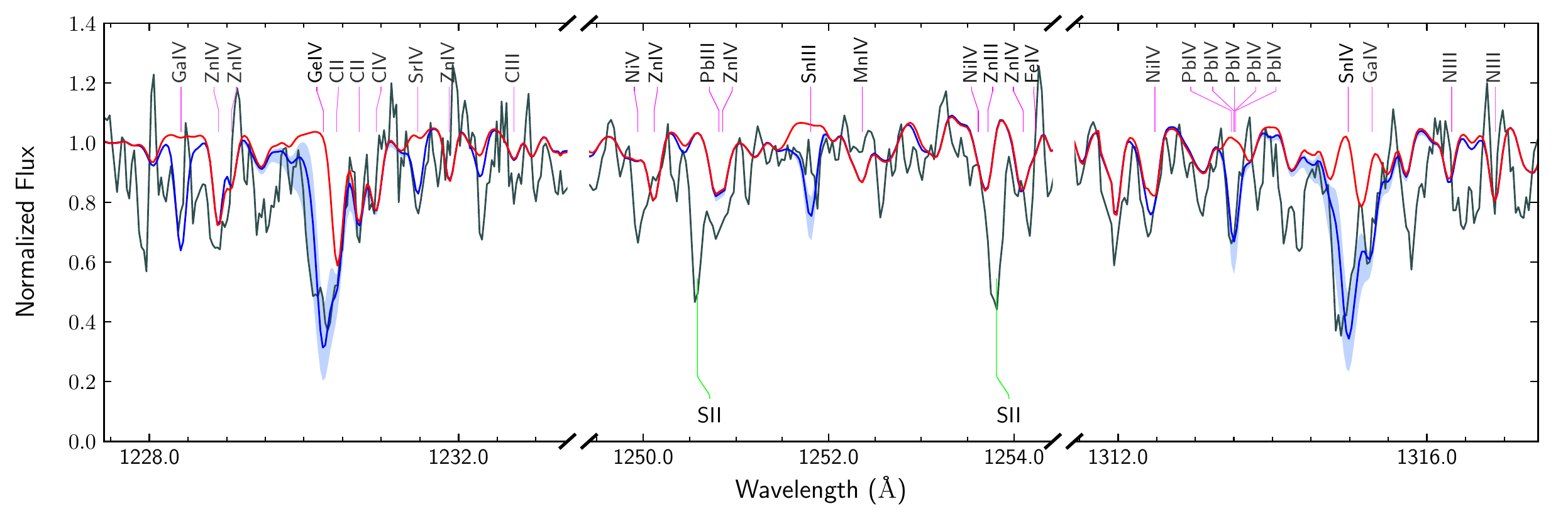}
	\caption{Comparison between the IUE and synthetic spectra in the ranges where Ge, Sn and Pb lines are seen. 
	The model spectra shown in blue includes the estimated abundances of Ge and Sn (from  Table~\ref{table_abund}) with the uncertainty range indicated by a shaded area. The lower bound of the shaded area for the Pb line represents the upper limit for Pb listed in Table~\ref{table_abund} plus its uncertainty (i.e. log Pb/H = $-$6.7). A synthetic spectrum computed without the contribution of Ge, Sn, and Pb is shown as comparison in red. All spectra are smoothed with a 3-pixel box filter for clarity.
	}
 	\label{spec_iue}
 \end{figure*}
 
\begin{table}
\small
\centering
\caption{Abundance results by number relative to hydrogen (log $N$(X)/$N$(H)), by number fraction (log $\epsilon$), and number fraction relative to solar (log $\epsilon$/$\epsilon_{\odot}$).
}
\label{table_abund}
\setstretch{1.4}
\begin{tabular}{l@{\hspace{2pt}}rrr}
\toprule
\toprule
 Element  & \multicolumn{1}{c}{log ($N_\mathrm{X}/N_{\mathrm{H}}$)} & \multicolumn{1}{c}{log ($\epsilon$)} & \multicolumn{1}{c}{log $\epsilon$/$\epsilon_{\odot}$} \\
\midrule
H       &     $ 0.0\phantom{0  \pm 0.05}$ &     $-0.16 \pm 0.02$ &                            $-0.12 \pm 0.02$ \\
He      &     $-0.36 \pm 0.05$ &     $-0.52 \pm 0.04$ &                             $0.59 \pm 0.04$ \\
C       &  $-2.94\pm0.27$ &          $-3.10\pm0.27$ &                                     $ 0.51\pm0.28$ \\
N       &  $-3.41\pm0.12$ &          $-3.57\pm0.12$ &                                     $ 0.63\pm0.13$ \\
O       &  $-4.09\pm0.23$ &          $-4.24\pm0.23$ &                                     $-0.90\pm0.24$ \\
Mg      &  <$-4.74^{+0.30}$ &          <$-4.90^{+0.30}$ &                                     <$-0.46^{+0.30}$ \\
Si      &  $-5.91\pm0.11$ &          $-6.07\pm0.11$ &                                     $-1.54\pm0.11$ \\
S       &  <$-5.70^{+0.30}_{}$ &  <$-5.86^{+0.30}_{}$ &                         <$-0.94^{+0.30}_{}$ \\

Ti      &  $-5.69\pm0.15$ &          $-5.85\pm0.15$ &                                     $ 1.24\pm0.16$ \\
Cr      &  $-5.69\pm0.17$ &          $-5.85\pm0.17$ &                                     $ 0.55\pm0.18$ \\
Mn      &  <$-5.30^{+0.40}_{}$ &  <$-5.46^{+0.40}_{}$ &                          <$1.15^{+0.40}_{}$ \\

Fe      &  $-4.65\pm0.14$ &          $-4.81\pm0.14$ &                                     $-0.27\pm0.15$ \\
Co      &  $-5.86\pm0.21$ &          $-6.02\pm0.21$ &                                     $ 1.03\pm0.23$ \\
Ni      &  $-4.54\pm0.11$ &          $-4.70\pm0.11$ &                                     $ 1.12\pm0.12$ \\
Zn      &  $-5.11\pm0.15$ &          $-5.27\pm0.15$ &                                     $ 2.21\pm0.16$ \\
Ga      &  $-5.20\pm0.40$ &          $-5.36\pm0.40$ &                                     $ 3.64\pm0.42$ \\
Ge      &  $-5.80\pm0.60$ &          $-5.96\pm0.60$ &                                     $ 2.43\pm0.63$ \\
Sr      &  $-4.40\pm0.40$ &          $-4.56\pm0.40$ &                                     $ 4.61\pm0.41$ \\
Y       &  $-5.00\pm0.40$ &          $-5.16\pm0.40$ &                                     $ 4.67\pm0.41$ \\
Zr      &  $-4.99\pm0.10$ &          $-5.14\pm0.10$ &                                     $ 4.31\pm0.11$ \\
Sn      &  $-6.00\pm0.60$ &          $-6.16\pm0.60$ &                                     $ 3.84\pm0.63$ \\
Pb      &  <$-7.30^{+0.60}_{}$ &  <$-7.46^{+0.60}_{}$ &                          <$2.83^{+0.60}_{}$ \\

\bottomrule
\end{tabular}
\end{table}

Iron and nickel show numerous lines over the GHRS ranges and their abundances were obtained by fitting multiple wavelength ranges containing mostly lines of iron or nickel. 
The estimated iron abundance is slightly below the solar value, while the nickel abundance is about ten times solar. 
With the iron and nickel abundances set to their best fit values, the same method was applied to derive the abundances of the iron-group elements Cr, Co, and Zn. 
The Mn lines are all relatively weak and strongly blended in the GHRS range.
We derived an upper limit of log~Mn/H < $-$5.7 using several blended lines (e.\,g.~Mn\,\textsc{iv}\,$\lambda\lambda$1340.66, 1341.93, 1442.83). 

The two \ion{Si}{iv} resonance lines indicate an abundance of log Si/H = $-$5.9 $\pm$ 0.1, meaning that this element is depleted in the star.
As for sulfur, we used the \ion{S}{v}\,$\lambda$1501.76 line to derive an upper limit of 1/10 solar. 
We also identified some strong Ti lines, most notably Ti\,\textsc{iv}\,$\lambda\lambda$1451.74, 1467.34, 1469.19, and Ti\,\textsc{iii}\,$\lambda$1455.20. 
Fitting these lines gives an abundance of log Ti/H = $-$5.69 $\pm$ 0.15, which is about 17 times solar.

The GHRS spectra cover several Ga\,\textsc{iv} lines (Ga\,\textsc{iv}\,$\lambda\lambda$1338.13, 1347.08, 1351.07, 1395.55, 1402.61), but also the very strong Ga\,\textsc{iii}\,$\lambda$1495.045 line.
We derived an abundance of log Ga/H = $-$5.2 $\pm$ 0.4 from these lines, about 4000 times solar. 
The Ga\,\textsc{iii}\,1507.96\,\AA\ line listed in \citet{morton03} was excluded from our models since it is not observed at the stated position and was also discrepant in the spectrum of HZ\,44 (Dorsch et al. 2019, submitted). 
The strongest Sr line in the GHRS spectra of \feige\ is by far Sr\,\textsc{iv}\,$\lambda$1331.13, although several weaker lines are also visible (e.\,g.~Sr\,\textsc{iv}\,$\lambda\lambda$ 1394.94, 1492.07).
A comparison with synthetic spectra is shown in Fig.\,\ref{ghrs_sr} and suggests an abundance of $\log$ Sr/H = $-$4.4 $\pm$ 0.4, which is more than 4 dex higher than the solar value. 

The Ge\,\textsc{iv}\,$\lambda$1229.84 resonance line is visible in the low S/N IUE spectrum of \feige\ (see Fig.~\ref{spec_iue}), while the GHRS data covers only the two blended Ge\,\textsc{iv} lines ($\lambda\lambda$ 1494.89, 1500.61). From these lines we derived log Ge/H = $-$5.8 $\pm$ 0.6.
The Sn\,\textsc{iv} resonance lines ($\lambda\lambda$\,1314.54, 1437.53) are outside of the GHRS ranges but are visible in the IUE spectrum (Fig.~\ref{spec_iue}). Based on these features, we estimated an abundance of log Sn/H = $-$6.0 $\pm$ 0.6. We note however that the predicted \ion{Sn}{iii} line at 1251.4 \AA\ does not appear in the observation. This is reminiscent of the issue discussed in \citet{lat16} where the authors found an important discrepancy between the abundances indicated by the \ion{Sn}{iii} line and the \ion{Sn}{iv} resonance doublet. The discrepancy observed by the authors, although in a colder sdB, was similar; the abundance indicated by the fit of the \ion{Sn}{iv} lines produced an \ion{Sn}{iii} line much stronger than observed. 

\begin{figure*}
\centering
\includegraphics[width=15cm]{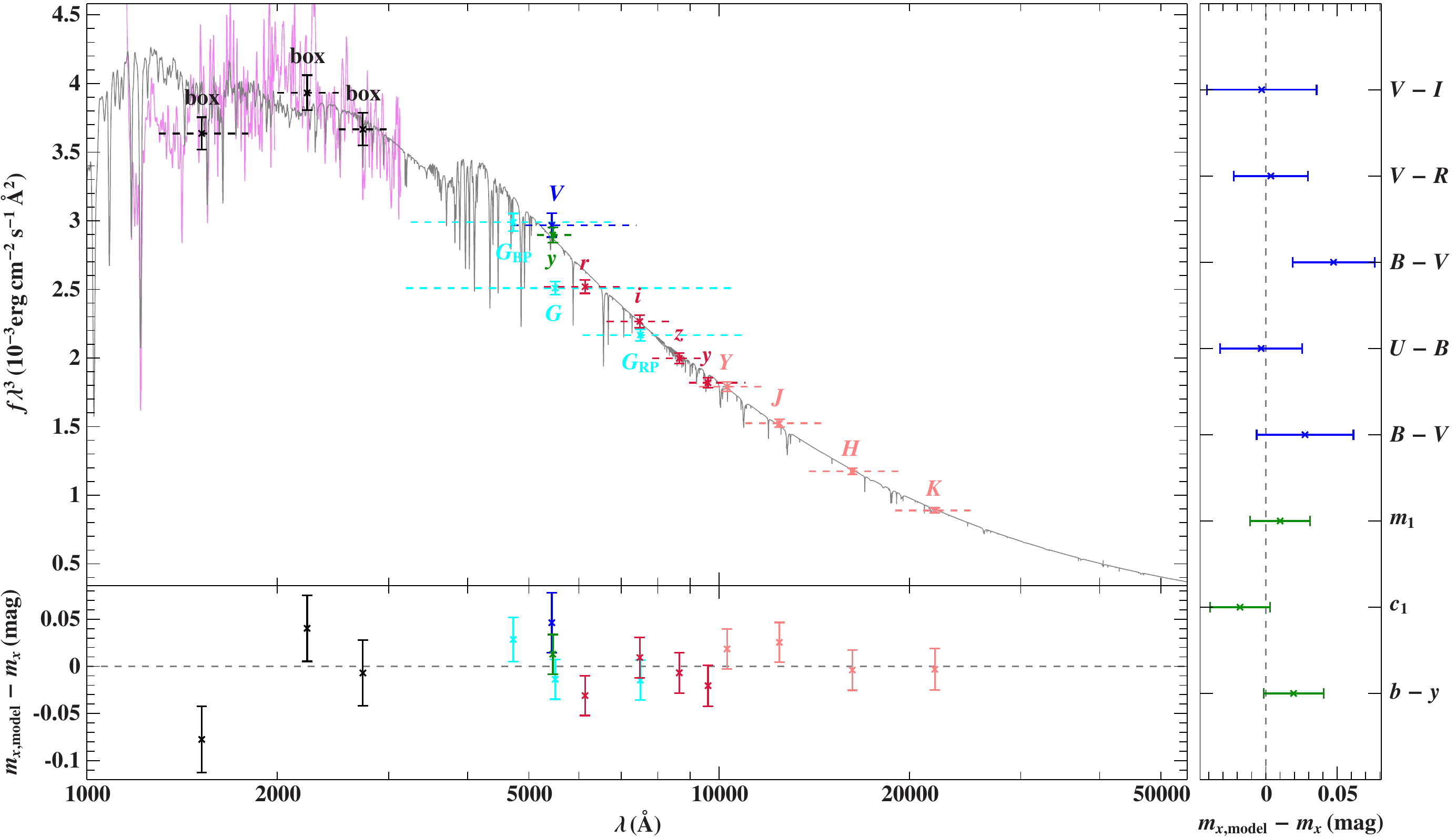}
\caption{Comparison of the synthetic spectrum (gray line) of \feige\ with photometric data.
The three black data points labeled ``box'' are binned fluxes from a low-dispersion IUE spectrum. 
Filter-averaged fluxes are shown as colored data points that were converted from observed magnitudes (the dashed horizontal lines indicate the respective filter widths). 
The residual panels at the bottom and right hand side respectively show the differences between synthetic and observed magnitudes/colors. 
The following color codes are used to identify the photometric systems: Johnson-Cousins (blue), Str\"omgren (green), \textit{Gaia} (cyan), UKIDSS (rose), and 2MASS (red).}
\label{fig:sed_fit}
\end{figure*}

A final element that is interesting in the context of heavy-metal enrichment is lead. Unfortunately, the GHRS ranges do not cover strong Pb lines, but the Pb\,\textsc{iv} line at 1313\,\AA\ in the IUE spectrum (Fig.~\ref{spec_iue}) excludes Pb abundances higher than log Pb/H  $\lesssim -7.3$ (700 times solar). Because the only lead line that we could identify is in the IUE spectrum and not particularly strong (compared to the \ion{Sn}{iv} resonance doublet) we prefer to state only an upper limit on the Pb abundance.  
We note that \citet{otoole04} reported the detection of Pb in the GHRS observation of \feige, although it is not mentioned which lines were detected in that star. Among the Pb lines investigated by the author, the transitions in the GHRS ranges unfortunately do not have oscillator strengths available.

In addition to the elements mentioned above, we also looked for lines of other chemical species that have been observed in hot subdwarfs, for example P, Al, Ar, and V. No convincing identification or abundances could be determined although in some cases we tentatively estimated an upper limit. 
We include in Appendix B a short discussion on these additional elements.

 \section{Spectral energy distribution and stellar mass}

Photometry, which is apparent magnitudes and colors, provides important information to constrain the angular diameter of a star. We used photometric measurements from the ultraviolet to the infrared as listed
 in Table \ref{tab:photometry} to construct the observed spectral energy distribution (SED) of \feige. For the theoretical SED, shown in gray in Fig.~\ref{fig:sed_fit}, we used our final synthetic spectrum of the star.  
Interstellar reddening was also taken into account by multiplying the synthetic flux with a reddening factor $10^{-0.4 A(\lambda)}$ using the extinction curve of  
\citet{Fitzpatrick1999} and assuming an extinction parameter $R_V = A(V) / E(B$--$V)$ = 3.15.

We derived the angular diameter ($\Theta$) and the interstellar reddening 
$E(B$--$V)$ by matching the observations 
with the synthetic SED and color indices obtained from the final model via $\chi^2$ minimization \citep[for details see][]{2018OAst...27...35H}.
The residuals of the SED and color fits are shown in Fig.~\ref{fig:sed_fit} along with the SED where the y-axis represents the flux density multiplied by the wavelength to the power of three (F$_\lambda\, \lambda^3$). Using this y-axis quantity reduces the otherwise steep slope of the SED on such a broad wavelength range. The angular diameter and interstellar reddening obtained from the fit are listed in Table~\ref{tab:SED}, as well as additional parameters that can be further derived: radius, mass and luminosity. 

\begin{table}
\setstretch{1.2}
\caption{Parallax and parameters derived from the SED fitting.}

\label{tab:SED}
\vspace{-8pt}
\begin{center}
\begin{tabular}{lc}
\toprule
\toprule
Quantity & Derived value  \\ 
\midrule
$\varpi$\,(mas)    &  $1.860  \pm 0.067$  \\
$d$\,(pc)		& $538 \pm 19$		\\
$\theta\,(10^{-11}\,\mathrm{rad})$	& $1.106 \pm 0.005$		\\
$E_{B-V}$		& $0.024 \pm 0.003$	\\
$R/R_\odot$		& $0.132 \pm 0.005$	\\
$M/M_\odot$		& $0.54 \pm 0.13$	\\
$L/L_\odot$ 	& $26.7 \pm 1.5$	\\
\bottomrule
\end{tabular}
\end{center}
\end{table}

 \begin{figure*}
 	\centering
 	\includegraphics[width=0.99\textwidth]{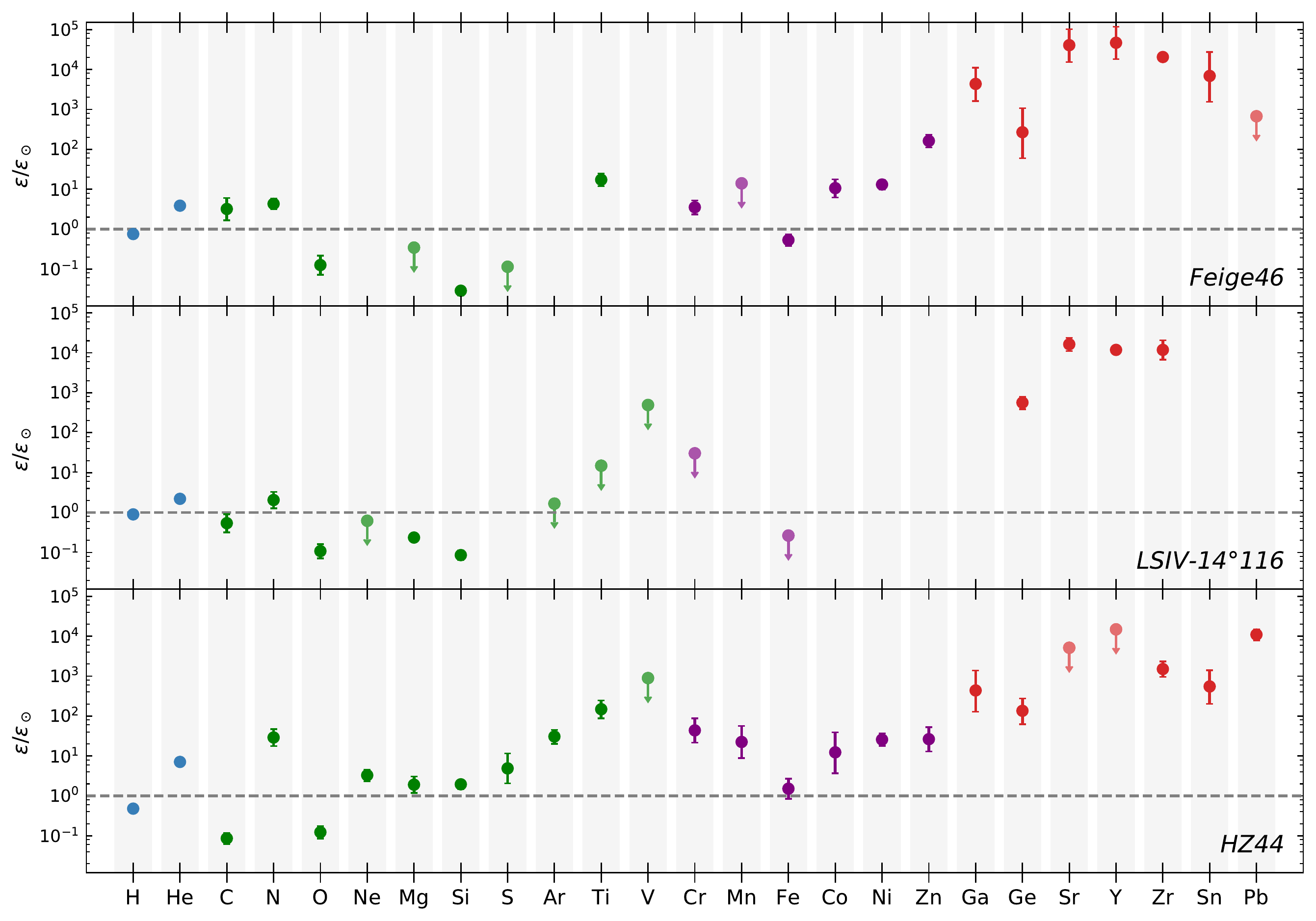}
	\caption{Comparison between the abundance pattern of \feige, \lsiv\ \citep{nas11} and HZ\,44 (Dorsch et al. 2019). Light elements ($23\leq\mathrm{Z}$) are marked by green symbols, iron-peak elements ($24\leq\mathrm{Z}\leq30$) in purple, and heavier elements ($\mathrm{Z}\geq31$) in red.}
 	\label{abund_comp}
 \end{figure*}

The stellar radius is obtained from the angular diameter and the parallax ($\varpi$)
\begin{equation}\label{eq:one}
  R = \Theta/(2\,\varpi). 
  \end{equation}
The stellar mass is derived from the surface gravity ($\log g = 5.93\pm0.10$), the angular diameter, and the parallax via 
  \begin{equation}\label{eq:two}
  M = g\,\Theta^2/(4\,G\,\varpi^2). 
  \end{equation}
Finally, the effective temperature ($T_{\rm eff} = 36100\pm 500$\,K) is also used to derive the luminosity from
\begin{equation}\label{eq:three}
  L = 4\pi\,\sigma\,(\Theta/2\,\varpi)^2\,T_{\rm eff}^4. 
\end{equation}
The stellar mass obtained is somewhat larger than, although consistent with, the canonical EHB mass for the core-helium flash (0.46\msun).

\section{Discussion}

The strong enrichment measured for Sr, Y, and Zr readily confirmed our initial suspicion that \feige\ belongs to the group of iHe subdwarfs that are also enriched in heavy metals. We summarized the chemical composition of \feige\ in Fig.~\ref{abund_comp} and included that of \lsiv\ as well as HZ\,44, another iHe subdwarf, as a comparison. 
The abundance pattern of Ge, Sr, Y, and Zr in the two pulsators is strikingly similar, with the latter three elements being similarly enhanced by more than 10~000$\times$ with respect to the solar abundances. The abundance pattern of C, N, O, Mg, and Si is also alike in both stars. Unfortunately, the limited wavelength range of the optical spectrum of \lsiv\ analyzed by \citet{nas11} did not provide much information on the abundance of elements between sulfur and gallium. We recall here that \feige\ is about 1000 K hotter than \lsiv\ and more He-enriched by $\sim$0.3 dex.
On the other hand, the abundances of HZ\,44 show clear differences with respect to the other two stars. Although HZ\,44 is also enriched in heavy metals, the abundances of Sr, Y, and Zr are not as large and Pb is the most enriched element in this star. In addition, the abundance pattern of the light elements is different: C and N follow the CNO cycle pattern, with carbon depletion and nitrogen enrichment, and the elements from Ne to Ar are slightly enriched with respect to solar while they appear to be mostly depleted in the two pulsators. The iron-group elements, however, follow a similar pattern in \feige\ and HZ\,44, and, beside iron, are enriched. HZ\,44 is hotter ($\sim$39 000 K) and more helium-rich (log He/H = 0.08) than the two pulsators. 

Few heavy-metal iHe subdwarfs have an extensive abundance portrait, which is partly due to the fact that deriving abundances for most of the iron-group elements requires UV spectroscopy. As seen for \feige, UV observations are also valuable to derive abundances of elements like Ga, Ge, Sr, Sn, and Pb. 
With incomplete chemical portraits for many of the heavy-metal subdwarfs, it is difficult to make proper comparisons. Nevertheless, \feige\ and \lsiv\ are so far the only iHe-sdOBs with such high overabundances ($\sim$3.0$-$4.5 dex) of Ge, Sr, Y, and Zr. At the moment, only one other star has known enrichment of the same order of magnitude in Y and Zr, but its abundance of Ge and Sr is unconstrained (HE 2359-2844, \citealt{nas13}). 

 \begin{figure*}[t]
 	\sidecaption
 	\includegraphics[width=12cm]{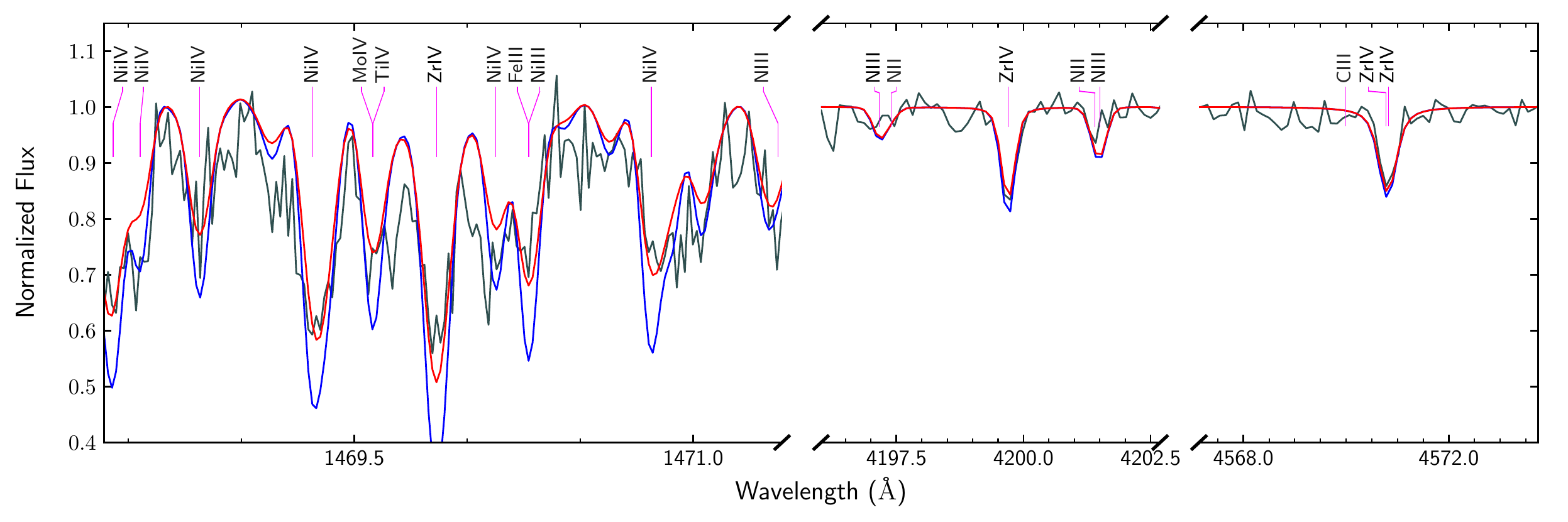}
	\caption{Comparison between the observed spectra and synthetic spectra having $v_{t}$ = 2~\kms, $v \sin i$ = 12 \kms\ (red) and $v_{t}$~=~5~\kms, $v \sin i$~=~0~\kms\ (blue).  }
 	\label{vtb}
 \end{figure*}
 
The spectral lines of \feige\ required an additional broadening, that we could reproduce by applying a rotational broadening of $v$ sin$i$ = 10 \kms\ to the synthetic spectra. However, this does not necessarily imply that the broadening observed is caused by stellar rotation. For one thing, the Fourier analysis of the star's light curve revealed peaks that could be associated with rotational splitting, indicating a rather long rotation period of $\sim$40 days \citep{lat19}. Furthermore, \citet{jeff15} showed that the atmospheric motion due to pulsations in \lsiv\ produces radial velocity variations with an amplitude of $\approx$ 5 \kms. It is thus possible that the pulsations in \feige\ are responsible for the extra broadening observed, as the exposure time of our different spectra are long enough (1500$-$5400\,s) to cover a significant fraction of the pulsation periods (2294$-$3400\,s). 
Broadening due to pulsations has also been reported in other sdBs \citep{kuass05,telt08,groo19}. The planned observations of \feige\ by the Transiting Exoplanet Survey Satellite \citep{ricker14} in sector 22 should provide with a longer (27 days) light-curve that will allow us to confirm the presence of rotational splitting and provide a better estimate of the rotation period.

An alternative mechanism that could produce line broadening is microturbulent velocity. For instance, in their analysis of \lsiv, \citet{nas11} adopted a microturbulent velocity of $v_t$ = 10 \kms\ based on six optical \ion{C}{iii} lines. However \citet{jeff17} found the optical \ion{N}{iii} lines in [CW83]\,0825+15, another iHe subdwarfs with overabundances of Y and Pb, to be rather insensitive to $v_t$ and adopted a value of $v_t$ = 2 \kms. To investigate the effect of microturbulence, we computed additional
synthetic spectra with $v_t$ = 5 \kms\ and no rotational broadening. We compared this spectrum with our current best model (including $v_t$ = 2 \kms) and the spectra of \feige\ around the \ion{Zr}{iv} lines in Fig.~\ref{vtb}. While the optical Zr lines are minimally affected by the change of microturbulent velocity, the stronger lines in the UV become significantly deeper with increasing microturbulence. While the Zr abundance derived from the optical lines (log Zr/H = $-$5.0) with rotational broadening reproduces well the Zr lines seen in the GHRS spectra, this is not the case when the microturbulence is increased. Following this comparison, increasing the microturbulent velocity does not appear as a suitable option to reproduce the observed line broadening.  

\section{Conclusion}

We performed an abundance analysis of the pulsating iHe-sdOB star \feige\ using optical and UV spectroscopic data. This allowed us to derive abundances for 17 metallic elements, including Ge, Sr, Y, Zr, and Sn.  Our results not only confirmed that the star belongs to the small class of ``heavy-metal'' hot subdwarfs, but also showed that \feige\ and \lsiv, the only two members of the V366 Aqr pulsating class, have a very similar abundance pattern that is characterized by extreme over-abundances of Sr, Y, and Zr ($\sim$4.0$-$4.5 dex). Among the handful of other known heavy-metal hot subdwarfs, none so far had abundances reaching the values reported in \lsiv\ for this trio of elements. We might wonder about the fact that the only star in which we found enhancement for these three elements as high as in \lsiv\ is also the one displaying the same type of pulsations. Could this be related to the pulsating nature of the stars, or to their halo kinematics? With so few abundances and upper limit constraints in other stars, we unfortunately cannot draw any conclusions yet. The only two stars with a complete set of values, and upper limits, for Sr, Y, Zr, as well as Ge and Pb are the two hotter and more He-rich stars (including HZ\,44) analyzed by Dorsch et al. (2019; submitted). In these two stars, the enrichment in Ge, Sr, Y, and Zr is not as large as in the pulsators, although their Pb abundance is higher than in \feige. With radiative levitation likely playing a role in supporting such high abundances of these elements in the line-forming region, the effective temperature of the stars would influence their photospheric abundances. The temperature region at which \feige\ and \lsiv\ are found ($\sim$35~500 $-$ 37~000~K) might well be optimal for the radiative support of Sr, Y, and Zr. On the other hand, it is not clear whether radiative levitation can build up such large overabundances from an initial solar-like amount and additional production of heavy elements, for example during the recent evolution of the star, might be required.

\begin{acknowledgements}
 We thank Andreas Irrgang and Simon Kreuzer for the development of the SED fitting tool.
 M.L.\ acknowledges funding from the Deutsche Forschungsgemeinschaft (grant DR 281/35-1). This research has made use of NASA's Astrophysics Data System. 
 Based on observations made with the NASA/ESA Hubble Space Telescope, obtained from the MAST archive (prop. ID GO5319, P.I. Heber) at the Space Telescope Science Institute. STScI is operated by the Association of Universities for Research in Astronomy, Inc. under NASA contract NAS 5-26555. 
 \end{acknowledgements}

\bibliographystyle{aa}


\begin{appendix}

\section{Partition functions}

For chemical elements with atomic number $Z\leq30$, the necessary data to include a line in LTE are already implemented in \texttt{Synspec} 51 for all relevant ionization stages. For heavier elements, however, only
the stages \textsc{i-ii-iii} are implemented.
In sdOB stars, many elements are more highly ionized, with the dominant stages being \textsc{iv-v}. Thus additional atomic data had to be supplemented to allow lines of heavy elements such as Ge, As, Zr, and Pb to be included.
Table \ref{tab:data:levels} lists the number of levels considered for each additional ion and their ionization energy ($\chi _I$). 
We used the non-standard \texttt{PFSPEC} subroutine in \textsc{Synspec} to compute partition functions for these ions as it was already implemented and used for C\,\textsc{vi}, N\,\textsc{vi-vii}, and O\,\textsc{vi-viii}.
Although the subroutine includes the possibility to account for level dissolution and electron shielding, we simply used the Boltzmann equation to compute the partition function ($U$) of the additional ions:
\begin{equation*}
\begin{split}
U_i &= g_i\cdot \exp (-E_i/k_BT)\\
U &= \sum _i U_i
\end{split}
\end{equation*}
For each level $i$, $g_i$ is the statistical weight, $E_i$ the energy of the level, and $k_B$ is the Boltzmann constant.

\begin{table*}
\caption{Number of new levels per ion added to \textsc{Synspec} and the corresponding ionization potentials.}
\vspace{-14pt}
\label{tab:data:levels}
\begin{center}
\begin{tabular}{l@{\hspace{9pt}}c@{\hspace{9pt}}r@{\hspace{9pt}}c@{\hspace{25pt}}l@{\hspace{9pt}}c@{\hspace{9pt}}r@{\hspace{9pt}}c@{\hspace{25pt}}l@{\hspace{9pt}}c@{\hspace{9pt}}r@{\hspace{9pt}}c@{\hspace{9pt}}}
\toprule
\toprule
 Ion	& $N_\mathrm{levels}$ & $\chi_I$\,(eV) & Source &  Ion	& $N_\mathrm{levels}$ & $\chi_I$\,(eV) & Source	& Ion	& $N_\mathrm{levels}$ & $\chi_I$\,(eV) & Source	\\  
\midrule
Ga\,\textsc{iv}		& 191			& 63.24		& 1	&     Zr\,\textsc{vi}		& \phantom{1}96		& 96.38		& 1	&         Te\,\textsc{vi}		& \phantom{11}8		& 69.10		& 1	\\ 
Ga\,\textsc{v}		& \phantom{1}91		& 86.01		& 1	&     Zr\,\textsc{vii}	&  \phantom{11}1	& 112.00	& 1	&                 Te\,\textsc{vii}	& \phantom{1}59		& 124.20	& 1	\\
Ga\,\textsc{vi}		& 157			& 112.70	& 1	&     Zr\,\textsc{viii}	&  \phantom{11}1	& 133.70	& 1	&                 Te\,\textsc{viii}	& \phantom{11}1		& 143.00	& 1	\\
Ga\,\textsc{vii}	& 180			& 140.80	& 1	&     Nb\,\textsc{iv}		& 182			& 37.61		& 1	&         Xe\,\textsc{iv}		& \phantom{1}94		& 45.00		& 1	\\
Ge\,\textsc{iv}		& \phantom{1}39		& 45.72		& 1	&     Nb\,\textsc{v}		& \phantom{1}30		& 50.57		& 1	&         Xe\,\textsc{v}		& \phantom{1}53		& 54.14		& 1	\\
Ge\,\textsc{v}		& 101			& 90.50		& 1	&     Nb\,\textsc{vi}		& 104			& 102.07		& 1	& Xe\,\textsc{vi}		& \phantom{1}72		& 66.70		& 1	\\
Ge\,\textsc{vi}		& 104			& 115.90	& 1	&     Nb\,\textsc{vii}	& \phantom{1}31		& 119.10	& 1	&                         Xe\,\textsc{vii}	& \phantom{1}67		& 91.60		& 1	\\
Ge\,\textsc{vii}	& 167			& 144.90	& 1	&     Nb\,\textsc{viii}	& \phantom{11}1		& 136.00	& 1	&                         Ba\,\textsc{iv}		& \phantom{1}31		& 47.00		& 1	\\
Ge\,\textsc{viii}	& \phantom{11}1		& 176.40	& 1	&     Mo\,\textsc{iv}		& \phantom{1}80		& 40.33		& 1	&         Ba\,\textsc{v}		& \phantom{1}51		& 58.00		& 1	\\
As\,\textsc{iv}		& \phantom{1}33		& 50.15		& 1	&     Mo\,\textsc{v}		& 257			& 54.42		& 1	&         Ba\,\textsc{vi}		& \phantom{1}49		& 71.00		& 1	\\
As\,\textsc{v}		& \phantom{11}8		& 62.77		& 1	&     Mo\,\textsc{vi}		& 112			& 68.83		& 1	&         Tl\,\textsc{iv}		& \phantom{1}43		& 50.72		& 1	\\
As\,\textsc{vi}		& \phantom{1}43		& 121.19	& 1	&     Mo\,\textsc{vii}	& \phantom{1}95		& 125.64	& 1	&                 Tl\,\textsc{v}		& \phantom{11}1		& 62.60		& 1	\\
As\,\textsc{vii}	& \phantom{1}49		& 147.00	& 1	&     Mo\,\textsc{viii}	& \phantom{1}76		& 143.60	& 1	&                 Tl\,\textsc{vi}		& \phantom{11}1		& 80.00		& 1	\\
Se\,\textsc{iv}		& \phantom{1}28		& 42.95		& 1	&     In\,\textsc{iv}		& \phantom{1}17		& 55.45		& 1	&         Pb\,\textsc{iv}		& 102	& 42.33		& 1			\\
Se\,\textsc{v}		& \phantom{1}14		& 68.30		& 1	&     In\,\textsc{v}		& \phantom{1}32		& 69.30		& 1	&         Pb\,\textsc{v}		& \phantom{1}44		& 69.00		& 1	\\
Kr\,\textsc{iv}		& \phantom{1}78		& 50.85		& 1	&     In\,\textsc{vi}		& \phantom{11}1		& 90.00		& 1	&         Pb\,\textsc{vi}		& \phantom{11}1		& 82.90		& 1	\\
Kr\,\textsc{v}		& \phantom{1}42		& 64.69		& 1	&     Sn\,\textsc{iv}		& \phantom{1}20		& 40.74		& 1	&         Bi\,\textsc{iv}		& \phantom{1}37		& 45.32		& 1	\\
Sr\,\textsc{iv}		& 254			& 56.28		& 1	&     Sn\,\textsc{v}		& \phantom{1}24		& 77.03		& 1	&         Bi\,\textsc{v}		& \phantom{1}14		& 56.00		& 1	\\
Sr\,\textsc{v}		& 143			& 70.70		& 1	&     Sn\,\textsc{vi}		& \phantom{1}29		& 94.00		& 1	&         Bi\,\textsc{vi}		& 114	& 88.00		& 1			\\
Sr\,\textsc{vi}		& \phantom{1}21		& 88.00		& 1	&     Sb\,\textsc{iv}		& \phantom{1}28		& 44.20		& 1	&         Th\,\textsc{iv}		& \phantom{1}24		& 28.65		& 2	\\
Y\,\textsc{iv}		& 129			& 60.61		& 1	&     Sb\,\textsc{v}		& \phantom{11}8		& 55.70		& 1	&         Th\,\textsc{v}		& \phantom{11}1		& 58.00		& 1	\\
Y\,\textsc{v}		& 113			& 75.35		& 1	&     Sb\,\textsc{vi}		& \phantom{1}59		& 99.51		& 1	&         Th\,\textsc{vi}		& \phantom{11}1		& 69.10		& 1	\\
Y\,\textsc{vi}		& \phantom{11}1		& 91.39		& 1	&     Sb\,\textsc{vii}	& \phantom{11}1		& 117.00	& 1	&                 Th\,\textsc{vii}	& \phantom{11}1		& 82.00		& 1	\\
Zr\,\textsc{iv}		& \phantom{1}34		& 34.42		& 1	&     Te\,\textsc{iv}		& \phantom{1}15		& 37.40		& 1	&          			& 		&			 &	\\
Zr\,\textsc{v}		& 101			& 80.35		& 1	&     Te\,\textsc{v}		& \phantom{1}44		& 58.70		& 1	&          			& 		&			 &	\\
\bottomrule
\end{tabular} 
\end{center}
\vspace{-14pt}
\tablefoot{
References: (1) NIST (\url{https://physics.nist.gov/PhysRefData/ASD/lines_form.html}), (2) \url{http://web2.lac.u-psud.fr/lac/Database/Tab-energy/} }
\end{table*}

\section{Additional chemical elements investigated}

We also looked for lines of Al, P, Ar, Ca, Sc, V, Kr, Nb, Mo, In, Te, and Xe and describe our results in the following.\\
Two \ion{Al}{iii} lines are predicted in the GHRS range, \ion{Al}{iii} 1379.67 \AA\ matches an unidentified line with log Al/H = $-$5.6 but the \ion{Al}{iii} line at 1348.13 \AA\ is too strong at such an abundance and is more consistent with log Al/H = $-$6.2. \\
Phosphorus abundances higher than log P/H = $-$5.4 are excluded by several lines (e.g.,~\ion{P}{iii} $\lambda\lambda$ 1334.81, 1344.33, 1344.85, \ion{P}{iv} 1467.43, 1484.51, 1487.79). The two \ion{P}{iii} lines at 1334.81 and 1344.33 \AA\ also exclude abundances higher than log P/H = $-$6.0. \\
No argon lines could be clearly identified in the GHRS range, but abundances higher than log Ar/H = $-$3.2 can be excluded.
The \ion{Ca}{iii} lines $\lambda\lambda$1461.89, 1463.34, 1484.87, and 1496.88 exclude abundances higher than log Ca/H = $-$4.0. \\
Vanadium at an abundance of V/H = $-$6.0 improves the GHRS fit in some blended regions, and the \ion{V}{iv} lines at 1355.13, 1403.61 \AA\ would match otherwise unidentified lines. This is not reliable enough for an abundance measurement, but an upper limit of V/H = $-$5.4 can be estimated as many V lines are too strong at this abundance. \\
We searched for niobium using \ion{Nb}{iv} oscillator strengths provided by \cite{Tauheed2005}.
Several unidentified weak lines in the GHRS spectrum (e.g., \ion{Nb}{iv} $\lambda\lambda$1445.74, 1473.44, 1508.72) can be reproduced with an abundance of log Nb/H = $-$5.9. 
Other \ion{Nb}{iv} lines such as $\lambda\lambda$1330.60, 1447.49, 1487.26, and 1517.46 favor a lower abundance of log Nb/H = $-$6.3. \\
A molybdenum abundance of log Mo/H = $-$6.2 improves the overall match between the observations and our best model. 
The Mo abundance could be higher since several lines of that element fit with abundances up to log Mo/H = $-$5.6 (e.g., \ion{Mo}{iv} 1455.86, 1456.32, 1458.04 \AA).
However, a few Mo lines are too strong when adopting log Mo/H = $-$5.6 (e.g., \ion{Mo}{iv} 1351.27, 1412.43, 1440.87 \AA). \\
There are three noteworthy \ion{In}{iii} lines in the GHRS range: $\lambda\lambda$1403.08, 1487.67, and 1494.11.
The 1403.08 \AA\ line fits a strong, unidentified line with log In/H = $-$5.4, which is excluded by the other two lines. This line is also blended with a weak \ion{Ni}{iii} line that could be underestimated in the model.
On the other hand, the \ion{In}{iii} 1487.67 would exclude abundances higher than log In/H = $-$6.5. \\
Finally, predicted scandium, krypton, tellurium and xenon lines are all very weak and no meaningful upper limit could be estimated.

\section{Additional material} 
\subsection{Full spectral comparisons}

\begin{figure*}
\centering
\includegraphics[width=24cm,angle =90,page=1]{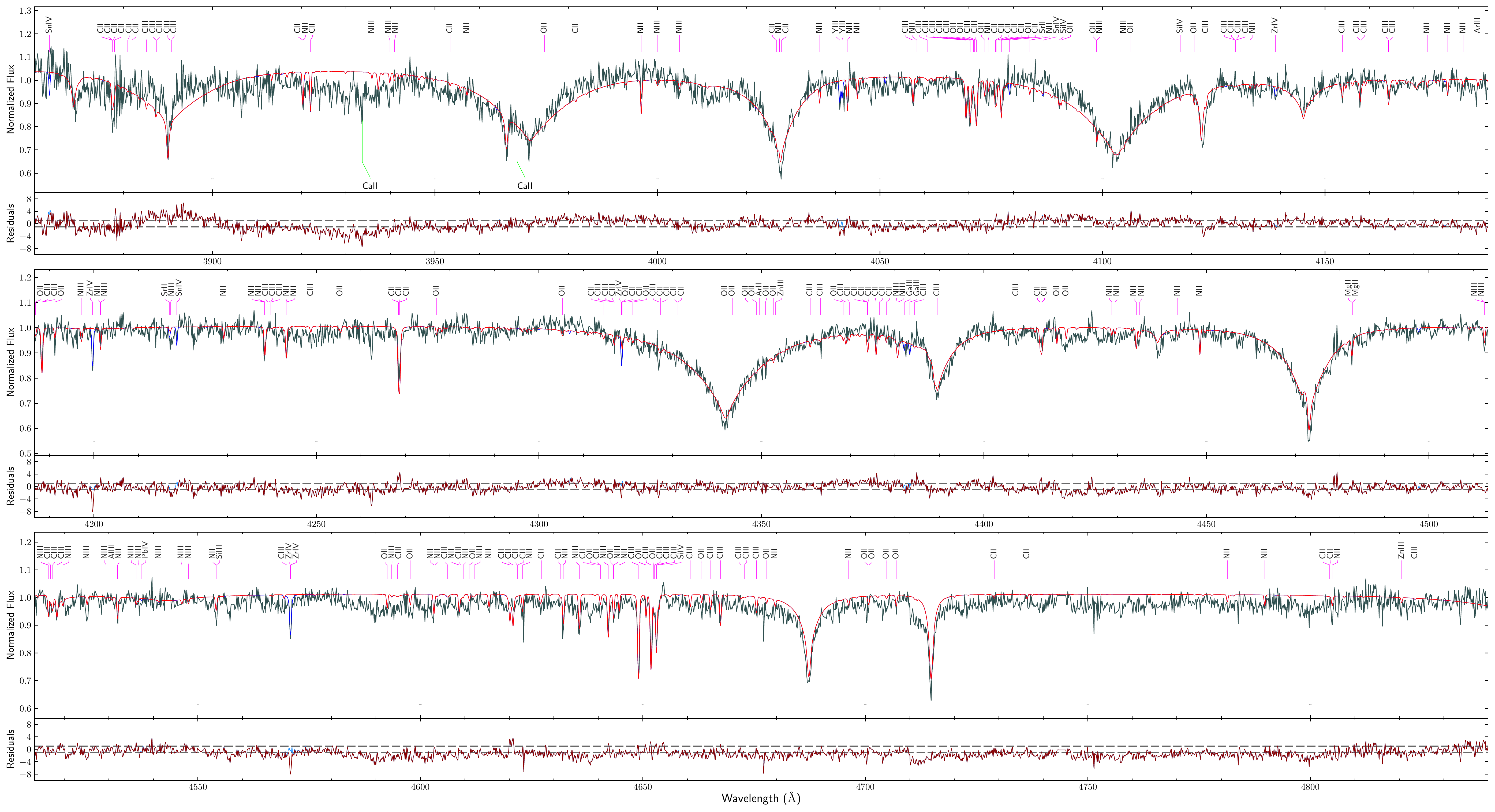}
\caption{CASPEC spectrum of \feige\ (gray) and the final model with $v_\mathrm{rot}\sin i =10$\,km\,s$^{-1}$ in red; the model in blue also includes lines from heavy elements (Z > 30). Note that mismatches of broad lines are due to shortcomings of normalisation procedure, which, however, do not compromise the measurement of sharp metal lines. }
\label{caspec_full}
\end{figure*}


\begin{figure*}
\centering
\includegraphics[width=24cm,angle =90,page=1]{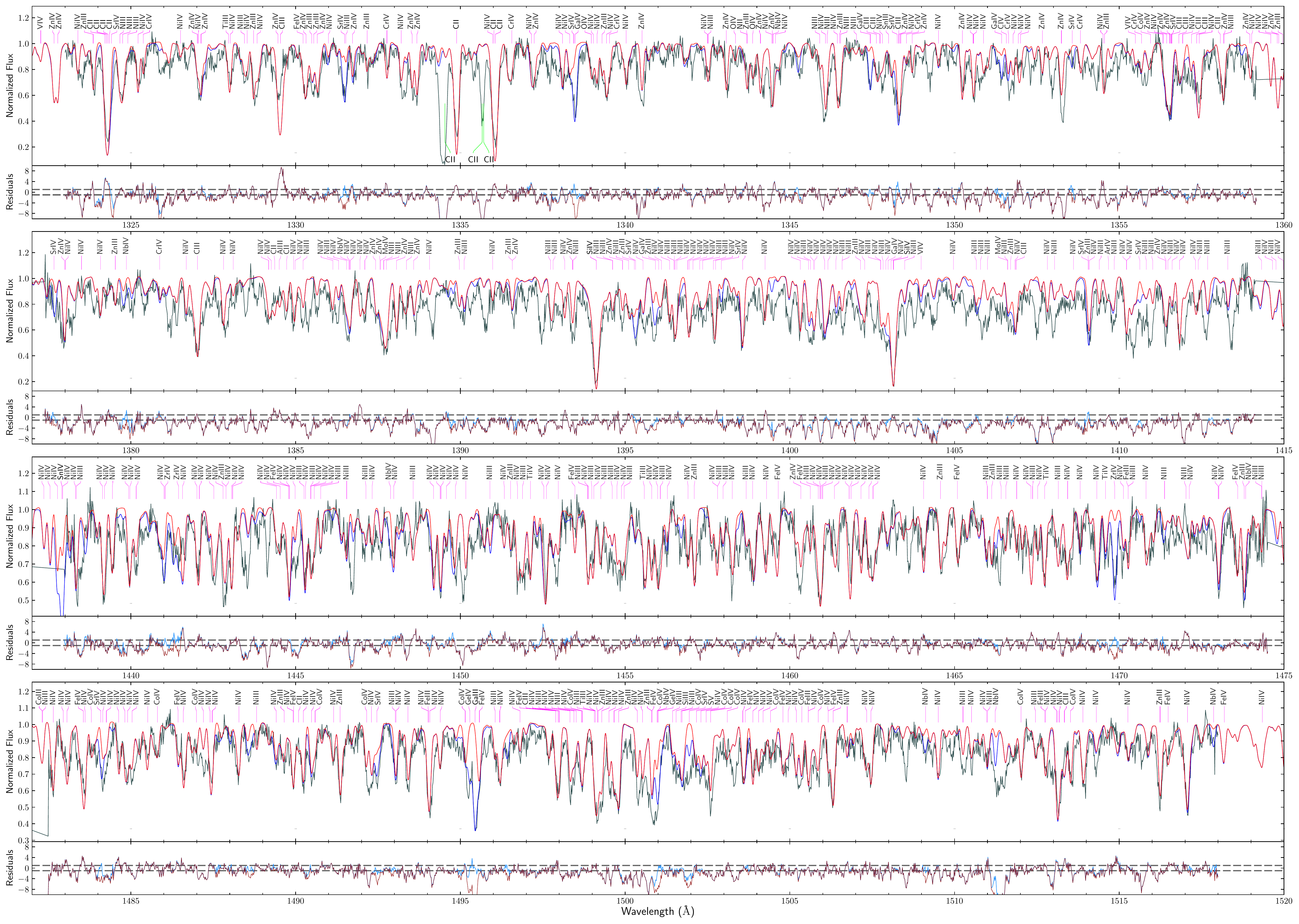}
\caption{GHRS spectrum of \feige\ (gray) and the final model with $v_\mathrm{rot}\sin i =10$\,km\,s$^{-1}$ in red; the model in blue also includes lines from heavy elements (Z > 30).}
\label{ghrs_full}
\end{figure*}

\begin{figure*}
\centering
\includegraphics[width=24cm,angle =90,page=1]{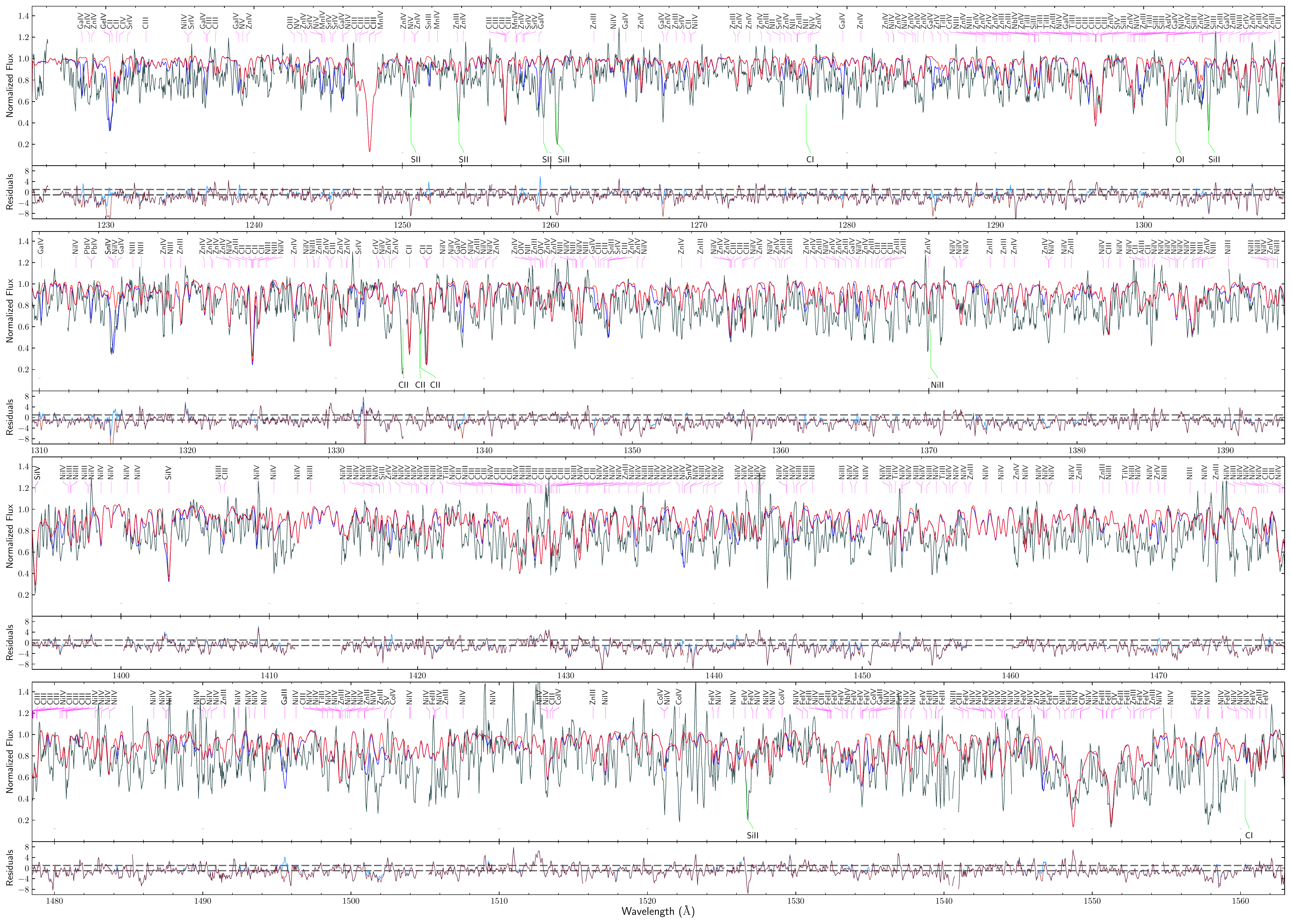}
\caption{Part of the IUE spectrum of \feige\ (gray) and the final model with $v_\mathrm{rot}\sin i =10$\,km\,s$^{-1}$ in red; the model in blue also includes lines from heavy elements (Z > 30). 
All spectra are smoothed with a 3-pixel box filter and the regions with bad data points in the IUE spectrum are not drawn.}
\label{iue_full}
\end{figure*}

\subsection{Photometric data}

\begin{table*}
\setstretch{1.2}
\small
\caption{Photometric data used for the SED-fit of \feige.}
\label{tab:photometry}
\vspace*{-12pt}
\begin{center}
\begin{tabular}{l l r r r c c}
\toprule
\toprule
    System & Passband & Magnitude & Uncertainty & Type & Reference\\ 
\midrule

Str\"omgren &        b$-$y  &  $-$0.129 &        &     color  & 
\cite{1970PASP...82.1305G}\\ 
Str\"omgren &         m1  &   0.074  &   &     color  & 
\cite{1970PASP...82.1305G}\\ 
Str\"omgren           & c1  & $-$0.202 &  &      color & 
\cite{1970PASP...82.1305G}\\
Str\"omgren &       y  &   13.3  &   &  magnitude  & 
\cite{1970PASP...82.1305G}\\ 
    UKIDSS &          H  &   14.084 & 0.003  & magnitude  & \cite[UKIDSS DR9: II/319/las9]{Lawrence2013_UKIDSS}\\ 
    UKIDSS &          J  &   13.950  & 0.002  & magnitude  & \cite[UKIDSS DR9: II/319/las9]{Lawrence2013_UKIDSS}\\  
    UKIDSS &          K  &   14.193  & 0.005  & magnitude  & \cite[UKIDSS DR9: II/319/las9]{Lawrence2013_UKIDSS}\\ 
    UKIDSS &          Y  &   13.869  & 0.002  & magnitude  & \cite[UKIDSS DR9: II/319/las9]{Lawrence2013_UKIDSS}\\ 
   IUE box & $1300-1800$\,\AA\  &    9.476  & 0.020  & magnitude  & \cite[VI/110/inescat]{Wamsteker2000_INES}\\ 
   IUE box & $2000-2500$\,\AA\  &    10.301  & 0.020  & magnitude  & \cite[VI/110/inescat]{Wamsteker2000_INES}\\ 
   IUE box & $2500-3000$\,\AA\  &    10.792  & 0.020  & magnitude  & \cite[VI/110/inescat]{Wamsteker2000_INES}\\ 
\textit{Gaia}       &     G       &   13.2512 & 0.001  & magnitude  & \cite[I/345/gaia2]{Gaia2018_VizieR}           \\  
\textit{Gaia}       &   GBP       &   12.9997 & 0.0074  & magnitude  & \cite[I/345/gaia2]{Gaia2018_VizieR}\\ 
\textit{Gaia}       &   GRP       &   13,5488 & 0.0021  & magnitude  & \cite[I/345/gaia2]{Gaia2018_VizieR}\\ 
Johnson    &     V$-$I     &   $-0.287$  & 0.023  &     color  & \cite{2001PASP..113..944W}\\ 
Johnson    &    V$-$R      &   $-0.129$  & 0.011  &     color  & \cite{2001PASP..113..944W}\\ 
Johnson    &     B$-$V     &   $-0.297$  & 0.014  &     color  & \cite{2001PASP..113..944W}\\ 
Johnson    &       V     &    13.26   & 0.017  & magnitude  & \cite{2001PASP..113..944W}\\ 
Johnson    &     B$-$V     &   $-0.277$  & 0.019  &     color  & \cite[II/168/ubvmeans]{Mermilliod2006_UVB}\\ 
Johnson    &     U$-$B     &   $-$0.114  & 0.014 & color  & \cite[II/168/ubvmeans]{Mermilliod2006_UVB}\\ 
PanSTARRS & i    &   13.885 & 0.003  & magnitude  & \cite[PanSTARRS DR1: II/349/ps1]{2016arXiv161205560C}\\ 
PanSTARRS & r    &   13.559 & 0.001  & magnitude  & \cite[PanSTARRS DR1: II/349/ps1]{2016arXiv161205560C}\\ 
PanSTARRS & y    &   14.384 & 0.004  & magnitude  & \cite[PanSTARRS DR1: II/349/ps1]{2016arXiv161205560C}\\ 
PanSTARRS & z    &   14.186 & 0.003  & magnitude  & \cite[PanSTARRS DR1: II/349/ps1]{2016arXiv161205560C}\\ 
\bottomrule 
\end{tabular}
\end{center}
\end{table*}

\end{appendix}

\end{document}